\documentstyle[aps,preprint]{revtex}
\begin{document}
%\preprint{ }
\draft

\title{\Large{Dynamic localization versus photon-assisted transport 
in semiconductor superlattices driven by dc-ac fields}}
 
\author{ W.-X. Yan$^{1,2}$, S.-Q. Bao$^{1,3}$, and  X.-G. Zhao$^{1,2,3}$}

\address{$^1$CCAST (World Laboratory) P.O. Box 8730, Beijing 100080, China\\$^2$Institute of Theoretical Physics, 
        Academia Sinica, P.O. Box 2735, Beijing 100080, China\\ 
   $^3$ Institute of Applied Physics and Computational Mathematics, \\P.O. Box 8009, Beijing 100088, China\\}
%\footnotesize

\maketitle 
\date{today} 
\tightenlines                        
\begin{abstract}
Via  the numerical analysis on the intraband dynamics of 
 optically excited semiconductor superlattices, we find that time-integrated
  squared THz emission signals can be used  for probing both dynamic
 localization and multi-photon resonance in the coherent regime.
 Competition effect between dynamic localization and photon-assisted transport
  has also been discussed.

\end{abstract}
                   
 \pacs{PACS numbers: 78.47.+p, 42.50.Md, 73.20.Dx, 78.20.Jq}

\widetext

 Recently, semiconductor superlattices (SL) driven by external electric
 fields have been a subject of intensive investigations both theoretically and
 experimentally.  The most significant example is Bloch oscillation
 (BO) which was theoretically predicted many years ago and experimentally 
 demonstrated recently in SL driven by an uniform electric  field.
 While, in the regime of tight-binding approximation, an intense laser field 
(ac fields) can make a charged particle localized around the initially 
 injected place, which is the so-called dynamic localization\cite{kenkre}. 
 Then the natural subsequent question will be the following: 
 What will happen if the electrons in SL is subject to the combined
 dc-ac fields?
 Actually, this question is answered by the recent several theoretical
 predictions and experimental observations. For example, multi-photon
 absorption\cite{monozon}, absolute negative conductivity\cite{keay},
 photon-assisted transport\cite{inverse}, fractional Wannier-Stark
 ladders\cite{fwsl}, etc.  Particularly, the recent experiment on the multiple
 quantum-well superlattices driven by dc-ac fields has confirmed through
 probing the static $I$-$V$ characteristics that the electrons in 
 SL can tunnel through the adjacent wells through the stimulated
 emission and absorption of photons\cite{inverse}. 
 In this facet, the common action of  
the dc-ac field plays the constructive role on the transport properties 
 in SL.  However, the dc-ac field can also play the destructive role on the 
transport in SL, which can be manifested in dynamic localization induced 
by dc-ac fields, provided that the ratio of the Stark frequency
 $\omega_B$($=eF_0d$, d is the SL lattice constant and $F_0$ is the amplitude
 of dc fields) to the ac field frequency $\omega$ is an integer $n$, and 
  $edF_1/\hbar\omega$ ($F_1$ is the amplitude of ac fields) is a root of
 ordinary Bessel function of the order $n$: $J_n(edF_1/\hbar\omega)=0$
\cite{kenkre}.
 In this case, the electrons will become localized.  Hence, the effect from both dc and ac fields play the dual role in the
 transport properties in SL.

 In the static regime, the usual method in the investigation of transport
 properties in SL under the influence of external fields is through the
 observation of static $I$-$V$ characteristics\cite{inverse}.  
 In the realm of ultra-short laser pulse (usually in hundreds of femto-second
 duration) generated coherent regime, what physical quantity can be used to
 probe the transport properties induced by the external fields?
 This question was partially answered by T. Meier et al with the help
 of the well-known
 semiconductor Bloch equations\cite{meier}. They found by monitoring the
 time-integrated squared THz emission signals that even in the presence of
 excitonic interactions comparable to the miniband widths, the carriers driven
 by pure intense laser fields (ac fields) reveal dynamic localization provided
 that ${edF_1}/{\hbar\omega}$ coincides with the roots of the ordinary
 Bessel function of order zero\cite{kenkre,meier}.
  Inspired by their work, we, in this report,
show by numerical simulations that in the dc-ac fields, the detection of 
 time-integrated squared THz signals not only can be used to probe 
 the dc-ac induced dynamic localization but also  be employed to   
 probe the photon-assisted transport in the presence of 
  excitonic interactions in the coherent regime.

We begin with the following semiconductor Bloch equations by including the 
longitudinal  external driving fields ${\bf F}(t)$:
\begin{equation}
\Bigl[\frac{\partial}{\partial t}-\frac{e}{\hbar}{\bf F}(t)\cdot\bigtriangledown_{\bf k}-\frac{i}{\hbar}[e_c({\bf k},t)-e_v({\bf k},t)]\Bigr]P({\bf k},t)
=\frac{i}{\hbar}[n_c({\bf k},t)-n_v({\bf k},t)]\Omega({\bf k},t)+\frac
{\partial P({\bf k},t)}{\partial t}|_{\rm coll}~,
\end{equation}
\begin{equation}
\Bigl[\frac{\partial}{\partial t}-\frac{e}{\hbar}{\bf F}(t)\cdot\bigtriangledown_{\bf k}\Bigr]n_{c,(v)}({\bf k},t)=\mp\frac{2}{\hbar}{\rm Im}[\Omega({\bf k},t)P^*({\bf k},t)]+\frac{\partial n_{c,(v)}({\bf k},t)}{\partial t}|_{\rm coll}~.
\end{equation}
In the above equations, $n_{c}({\bf k},t)$ and $n_{v}({\bf k},t)$ are the 
 populations in the conduction and valence bands respectively.
  $P({\bf k},t)$ is the interband polarization.
  $e_{c,(v)}({\bf {k}},t)= \epsilon_{c,(v)}({\bf k},t)-\sum\limits_{\bf k'}
V({\bf k, k'})n_{c,(v)}({\bf k'},t)$ are the electron and hole energies 
by taking into account of the excitonic interaction, while, $\epsilon_{c,(v)}
 ({\bf k},t)$ are the corresponding energies of the conduction
 and (valence) bands with no Coulomb interaction.  
$\Omega({\bf k},t)=\mu E(t)+\sum\limits_{\bf k'}V({\bf k,k'})P({\bf k'},t)$ 
is the renormalized Rabi frequency and $V({\bf k,k'})$ is the Coulomb
 potential in the quasi-momentum space. $E(t)$ is the optical field and 
assumed to be the Gaussian laser pulses\cite{meier}. ${\bf F}(t)$ is the 
combined dc-ac fields: ${\bf F}(t)={\bf F}_0+{\bf F}_1\cos(\omega t)$.
  Incoherent dissipative processes  are phenomenologically described by the
 last terms in Eqs. (1) and (2).

 If the energy quanta of the optical field $E(t)$ is off resonant with the 
 excitonic resonance,  the low excitation regime can be produced.
 In this regime,  the perturbation expansion of Eqs. (1) and (2)
 according  to the optical field can be performed by  the  following 
 procedure \cite{meier,shen}: $n_{c,(v)}({\bf k},t)=n_{c,(v)}^{(0)}
 ({\bf k},t)+n_{c,(v)}^{(2)}({\bf k},t)+\cdot\cdot\cdot;$ 
  $P({\bf k},t)=P^{(1)}({\bf k},t)+P^{(3)}({\bf k},t)+\cdot\cdot\cdot$. 
  The usual initial condition that  SL is in the ground state before the 
 optical pulse is triggered is adopted, i.e.,  
  $n_{v}({\bf k},t=0)=1.0$, $n_{c}({\bf k},t=0)=0$.
 Up to the second-order in the optical field, we can get the partial
 differential equations for both the first-order interband polarization 
 $P^{(1)}({\bf k},t)$ and the second-order electron (hole)  
 population density $n^{(2)}_{c,(v)}({\bf k},t)$\cite{meier}.

 For simplicity, we focus on the 1-D case, which is just the SL growth
 direction (assumed to be $z$ direction). The driving dc-ac field is assumed
 to be along this direction, and we adopt the contact Coulomb potential
 $V\delta(z-z')$, which carries the most important excitonic
 characteristic\cite{miller}. Under the above assumptions, the partial
 differential equations for the first order $P^{(1)}(k,t)$ and the 
 second order $n^{(2)}(k,t)$\cite{meier}  can be reduced to the following
 integro-differential equations with the help of the accelerated basis
 in the quasi-momentum $k$: ( by changing $k$ to $k-\eta(t)$, and
 $\eta(t)$ is defined as: 
 $\eta(t)=\frac{e}{\hbar}\int_0^tF(t') dt')$\cite{quade}.
\begin{eqnarray}
&&\Bigl[\frac{\partial}{\partial t}-\frac{i}{\hbar}
[\epsilon_c(k-\eta(t),t)-\epsilon_v(k-\eta(t),t)+V]\Bigr]{\tilde P^{(1)}}(k,t)\nonumber\\
& = & -\frac{i}{\hbar}\Bigl[\mu E(t)+V\sum\limits_q {\tilde P^{(1)}}(q,t)\Bigr] - \frac {{\tilde P^{(1)}}(k,t)}{T_1}~,
\end{eqnarray}

\begin{equation}
\frac{\partial {\tilde n^{(2)}}_{c,(v)}(k,t)}{\partial t}
=\mp\frac{2}{\hbar}{\rm{Im}}\Bigl[\Bigl(\mu E(t)+V\sum\limits_q {\tilde P^{(1)}}(q,t)\Bigr){\tilde P^{(1)*}}(k,t)\Bigr]-\frac{{\tilde n^{(2)}}_{c,(v)}(k,t)}{T_2}~,
\end{equation}
 where we have introduced the following new notations for convenience:
\begin{equation}
$${\tilde P^{(1)}}(k,t)=P^{(1)}(k-\eta(t),t),~~
 {\tilde n^{(2)}}_{c,(v)}(k,t)=n^{(2)}_{c,(v)}(k-\eta(t),t)~.
\end{equation}
 In Eqs. (3) and (4), we adopt the relaxation approximation by introducing
 the transverse time $T_1$ and longitudinal relaxation time $T_2$ respectively 
 to replace the collision terms.
 More sophisticated description of these dephasing processes calls for Monte
 Carlo simulation\cite{mc}.
 
As we have mentioned previously, what we want to do is to calculate the 
 time-integrated squared THz signal.  The THz emission signal in SL can be 
 expressed as $S_{\rm {THz}}(t)\propto \partial_t j^{(2)}(t)$, where
  $j^{(2)}(t)=\frac{e}{\hbar}\sum\limits_{i}\int{\partial \epsilon_i(k)}/
{\partial k}~n^{(2)}_i(k,t) dk~~ (i=c,v)$ 
 is the  current due to the non-equilibrium distribution of  
  electrons and holes excited by the Gaussian laser pulses.
 This current can be rewritten in the following equivalent form with the help
 of accelerated basis
 \begin{equation}
j^{(2)}(t)=\frac{e}{\hbar}\sum\limits_{i}\int\frac{\partial
 \epsilon_i(k-\eta(t))}{\partial k}~{{\tilde n^{(2)}}}_i(k,t) dk~~(i=c,v)~.
\end{equation}
Solving Eqs. (3) and (4)  in the accelerated  basis in $k$ space is more
 convenient than the direct integration in the usual quasi-momentum
 space.  In the following numerical simulation, we use the similar parameters
 as those used by T. Meier {\it et al}. 
 The combined miniband width $\Delta=\Delta_c+\Delta_v$
 =20 meV (here, we use the tight-binding model as in Ref.[6]); Coulomb potential
 strength $V$=10 meV.
  The  central frequency of the optical field is assumed
 to be located at 2meV below excitonic resonance which satisfies the
 off-resonant condition\cite{ex}.  The full width at half maximum
 of Gaussian laser pulse envelope $|E(t)|^2$ is chosen to be 100 fs.  Both
 transverse and longitudinal relaxation time $T_1$ and $T_2$ have been set to
 be 2 ps.  The energy quanta of the ac field $\hbar\omega$ is fixed to be
 20 meV.  To capture the information on dynamic localization, we set the
 ratio of the Stark frequency $\omega_B$ to the ac field frequency $\omega$
 to be an integer, e.g, $n=1$, without loss of generality. We change the ratio
 $eF_1d/\hbar\omega$  continuously, and  monitor the time-integrated squared 
 THz emission signal through the calculation of the following quantity:  
\begin{equation}
  T_s\propto\int d\tau |S_{\rm {THz}}(\tau)|^{2}~. 
\end{equation}
 The plot of the time-integrated THz signal vs. the ratio $eF_1d/\hbar\omega$
 has been shown in Fig.1.
 In this figure, it can be clearly seen that when the ratio $eF_1d/\hbar\omega$ 
 scans through 3.95, 6.95, 10.0, which are approximately the roots of the
 ordinary Bessel function of the first order $J_1$, the time-integrated
 squared signal falls into valleys. 
 While the peaks in the plot lies in the vicinity at those values of 
 $eF_1d/\hbar\omega$ that make $|J_1(eF_1d/\hbar\omega)|^2$ reach the 
 local maxima.
 From the above effect, it is obvious that $|J_1(eF_1d/\hbar\omega)|^2$
 reflects the signal $T_s$ profile  qualitatively.
 All the above phenomena demonstrate that
 even in the presence of Coulomb excitonic interaction whose strength is
 comparable to the miniband widths of SL, dc-ac induced dynamic localization
 still appears. This kind of dynamic localization can be probed through the
 observation of the signal $T_s$ just like the case of a pure ac
 field\cite{meier}.

  Another aspect of the combined dc-ac fields on the transport of the SL lies in that electrons/holes can tunnel through the adjacent wells through the photon 
 emission and absorption, when the ratio of the Stark frequency of dc field
 $\omega_B$ to the ac field frequency $\omega$ is an
 integer\cite{keay}. This phenomenon is  the 
 so-called {\it inverse Bloch oscillators}, which was found in a recent
 experimental study by Unterrainer et al in the static regime\cite{inverse}. 
 The coherent regime counterpart of this effect can also be found in our
 numerical calculation, which was shown in Fig.2. In this figure, we give
 the plot of the time-integrated 
 squared THz emission signal vs. the ratio $\omega_B/\omega$,
 and set the ratio $eF_1d/\hbar\omega$ to be 2.0.
 From the figure, we can see that the large peaks appear at the location
 where the ratio $\omega_B/\omega$ is an integer.
  The other distinctive peaks, which appear at fractional $\omega_B/\omega$
 can be attributed to the dynamical fractional Wannier-Stark ladders
 theoretically proposed recently\cite{fwsl,cold}.
  
  Let us look at the competition effect between the dc-ac induced dynamic localization and the photon-assisted transport facilitated by coupling of 
 the combined dc-ac fields. The competition  between these two effects 
 is shown in Fig.3, where we show the plot of the time-integrated 
 squared THz emission signals vs. the ratio $\omega_B/\omega$.
 In this figure, we set the ratio $eF_1d/\hbar\omega$ to be around 3.8
 which is the first root of the ordinary Bessel function of the first order.
 It can be clearly seen that the signal shows a deep valley when the ratio
 $\omega_B/\omega$ reaches around the unity. While, the peaks appearing 
 at $\omega_B/\omega=2,3$ are still present. In other words, the one-photon
 assisted transport has been suppressed by this specific selection of
  field parameters, and the dynamic localization prevails over the
 one-photon assisted transport.  If we shift the ratio from those values
 which are in the vicinity of  the roots of  the first-order ordinary Bessel
 function, 
the one-photon assisted transport  can revive. This regeneration can be
 clearly seen from the Fig.2. We can further suppress the two-photon
 resonance by selecting the ratio $\omega_B/\omega$ to be $2$, and 
 another ratio $eF_1d/\hbar\omega$ to be the roots of the ordinary Bessel
  function of the order $2$, i.e., make $J_2(eF_1d/\hbar\omega)=0$, which 
 we don't show for saving space. 
 
In summary, using the tight-binding model driven by dc-ac fields and  
 perturbatively solving the semiconductor Bloch equations in the accelerated 
 quasi-momentum space, we found that the time-integrated squared THz emission
 signal reveals both the dynamic localization and the multi-photon resonance
 in the coherent regime.
 The dynamic localization can be fulfilled when the ratio of 
 Stark frequency $\omega_B$ of dc fields  to the ac field frequency 
 $\omega$ is an integer $n$, and the another ratio $eF_1d/\hbar\omega$ 
 is located around one of the roots of the $n$-th order ordinary 
 Bessel function $J_n$.
 The dynamic localization is embodied  by suppressing the oscillation
 amplitude of the THz emission signals.
  Multiphoton assisted transport (resonance) can be identified 
 when time-integrated squared THz emission signals reach peaks 
 provided that  the ratio $\omega_B/\omega$ is an integer, and 
 another ratio $eF_1d/\hbar\omega$ is not in the vicinity of the roots of the 
 ordinary Bessel function of integral order.
 Otherwise, the photon-assisted transport will be destroyed by the 
 dynamic localization. Experimentally,  all the above findings can be probed 
 by detecting the time-integrated squared THz emission signals 
 from the ultra-short laser pulse excited SL driven
 by combined dc-ac fields. For example, one can use the similar expremental 
 arrangement by the UCSB group\cite{inverse}.

\begin{center}

 {\bf ACKNOWLEDGMENT}

\end{center}

 The authors thank Dr. T. Meier, Prof. W.-M. Zheng
 and Prof. J. Liu for the stimulating and useful discussions. This work was
 supported in part by the National Natural Science Foundations of China under
 grant No. 19724517, a grant of China Academy of Engineering and Physics, and
 China Postdoctoral Science foundation.

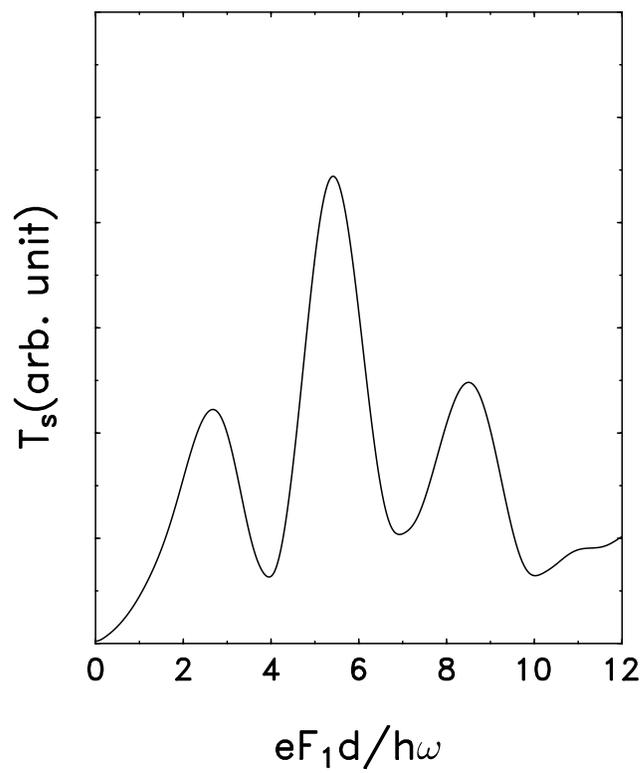
\begin{figure}

\caption{  The plot of time-integrated squared THz emission signal vs. 
 the ratio $eF_1d/\hbar\omega$, showing the  dynamic localization induced by
 dc-ac fields. The parameters used in this figure are declared in the text.}

\end{figure}

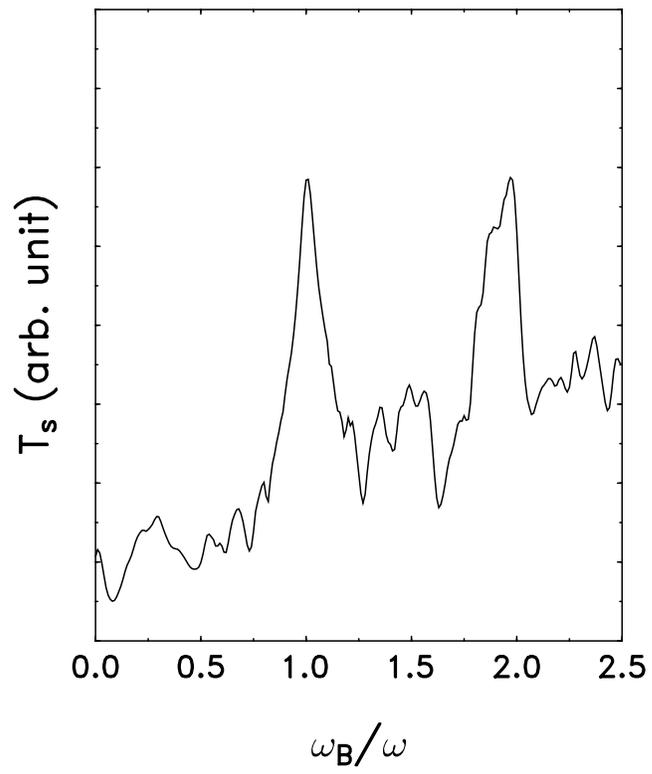
\begin{figure}

\caption{ The plot of time-integrated squared THz emission signal vs.
 the ratio $\omega_B/\omega$, multi-photon resonance can be clearly seen.
 We choose the ratio $eF_1d/\hbar\omega$ to be 2.0, which is well away from   
 the roots of the first-order ordinary Bessel function.} 

\end{figure}
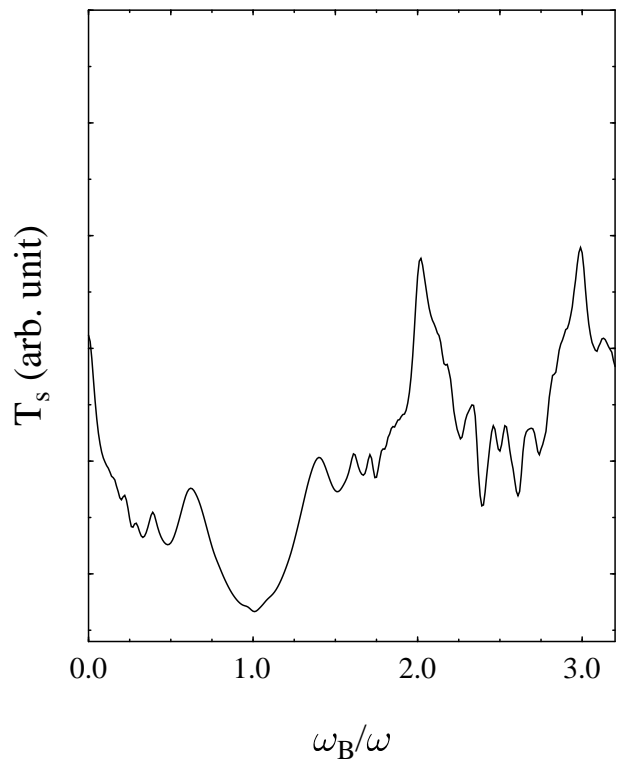
\begin{figure}

\caption{The  same plot as that of Fig.2, but we choose the ratio 
 $eF_1d/\hbar\omega$ to be 3.8, which is around the root of the first-order 
 ordinary Bessel functions.}

\end{figure}

\end{document}